\documentclass[aps,prb,twocolumn,showpacs,superscriptaddress]{revtex4-1}
\pdfoutput=1
\usepackage[titletoc,toc,title]{appendix}
\usepackage{amsmath,amssymb}
\usepackage{graphicx}
\usepackage{color}
\usepackage{rotating}
\usepackage{bm}
\usepackage{setspace}
\usepackage{hyperref}
\usepackage{float}
\usepackage{soul}
\usepackage{color}
\usepackage{braket}

\hypersetup{
colorlinks=true,
urlcolor= blue,
citecolor=blue,
linkcolor= blue,
bookmarks=true,
bookmarksopen=false,
}

\renewcommand{\vec}[1]{\bm{#1}}
\newcommand{\kp}{$\vec{k} \cdot \vec{p}$ }

\begin{document}

\preprint{APS/123-QED}

\title{Derivation of Wannier orbitals and minimal-basis tight-binding hamiltonians for twisted bilayer graphene: a first-principles approach}
\author{Stephen Carr}
\affiliation{Department of Physics, Harvard University, Cambridge, Massachusetts 02138, USA}
\author{Shiang Fang}
\affiliation{Department of Physics, Harvard University, Cambridge, Massachusetts 02138, USA}
\author{Hoi Chun Po}
\affiliation{Department of Physics, Massachusetts Institute of Technology, Cambridge, MA 02139, USA}
\author{Ashvin Vishwanath}
\affiliation{Department of Physics, Harvard University, Cambridge, Massachusetts 02138, USA}
\author{Efthimios Kaxiras}
\affiliation{Department of Physics, Harvard University, Cambridge, Massachusetts 02138, USA}
\affiliation{John A. Paulson School of Engineering and Applied Sciences, Harvard University, Cambridge, Massachusetts 02138, USA}
\date{\today}

\begin{abstract}
Twisted bilayer graphene (TBLG) has emerged as an important 
platform for studying correlated phenomena, 
including unconventional superconductivity, in two-dimensional systems. 
The complexity of the atomic-scale structures in TBLG has made even 
the study of single-particle physics at low energies around the Fermi level, quite challenging.
Our goal here is to provide 
 a convenient and physically motivated picture of single-particle physics in TBLG
using reduced models with the smallest possible number of localized orbitals.
The reduced models exactly reproduce the low-energy bands 
of \textit{ab-initio} tight-binding models, including the effects of atomic relaxations. Furthermore, we obtain for the first time the corresponding Wannier orbitals 
that incorporate all symmetries of TBLG, which are also calculated as a function of angle, a requisite first step towards incorporating electron interaction effects.
We construct eight-band and five-band models for the low-energy states 
for twist angles between $1.3^\circ$ and $0.6^\circ$.
The models are created using a multi-step Wannier projection technique 
starting with appropriate {\it ab initio} \kp continuum hamiltonians.
Our procedure can also readily  capture the perturbative effects of substrates
and external displacement fields while offering a significant reduction in complexity 
for studying electron-electron correlation  phenomena in realistic situations.
\end{abstract}

\maketitle

\section{Introduction}
Recent experimental advances in 
nanotechnology fabrication methods \cite{Riccardo2018} have made it possible to 
stack two or more layers of two-dimensional (2D) materials with exquisite
control of the angle of twist (denoted here as $\theta$) 
in the relative orientation of the layers.
This level of control in relative orientation, demonstrated in detail in 
twisted bilayer graphene (TBLG) but also applicable to other 2D layered systems, 
has opened new possibilities in controlling their electronic, transport, 
optical and other properties\cite{Li2009, Brihuega2012, Luican2011, Wong2015}
-- in essence introducing the new field of ``Twistronics''
\cite{Carr2017}. 
One of the most well known of twist-angle dependent phenomena 
is the emergence of correlated insulators and 
unconventional superconductivity in TBLG at $\theta \approx 1^\circ$, 
with a superconducting $T_c$ of a few degrees Kelvin 
\cite{Cao2018mott, Cao2018sc, Yankowitz2019}.
Theoretical studies of TBLG  near the critical (often referred to as ``magic'') angle of  
$\sim 1^\circ$ predicted uncommonly flat bands near the Fermi energy, 
providing some tantalizing hints for why strongly correlated phases can arise near that angle.
This single-particle band structure for TBLG has been modeled successful by density functional theory \cite{Trambly2010, Uchida2014}, tight-binding hamiltonians \cite{Morell2010, Carr2017, Nam2017, Angeli2018}, and continuum models \cite{Mele2010,Bistritzer2011,San-Jose2012, Weckbecker2016}.

All such models feature a large number of degrees of freedom, which render the study of electrons' correlation unfeasible.
Even the simplest, a continuum \kp model, requires a basis of hundreds of Bloch states, 
which is numerically intractable for proper many-body calculations.
A number of more simplified $N$-band models  have been proposed, 
where $N$ is typically in the range $2 - 10$, not taking into account 
the spin and $k$-space ``valley'' degrees of freedom, the latter referring 
to the point in reciprocal space where the band extrema near the Fermi level appear \cite{Ashvin_10band,Liang_Wan,Luijun_tBLG,Adrian_Mott,Liang_tBLG,TBLG_wan,arxiv1905.01887}.
These models are generally empirical, and chosen to address the flat bands near the magic-angle while taking advantage of the roughly $30$ meV band gaps that separate the flat bands 
and the rest of the energy bands with the inclusion of atomic relaxation \cite{Nam2017,Carr2018pressure}.
This assumes that the correlations responsible for the observed superconductivity can be accurately described with just the flat bands.
If the interaction strength in magic angle TBLG is larger than the gaps that separate the flat band manifold, a model which includes additional nearby bands is required.

Here we combine the accuracy of existing \textit{ab initio} models with the simplicity of a reduced model by directly constructing Wannier functions for a large range of twist angles around $\theta = 1^\circ$.
Our starting point is a microscopic \kp model for TBLG
which includes atomic relaxations \cite{Carr_10band,Carr_GSFE}, 
from which we derive three single-valley tight-binding models.
The first, and more comprehensive, is an $8$-band model, 
which includes the 2 flat bands and 3 bands above and 3 below them.
Two additional reduced models include $5$ bands, 
and involve the 2 flat bands and the nearest three bands either above 
or below them.
The two $5$-band models are only reliable when the flat band manifold is isolated in the band structure.
Near $0.85^\circ$, the gaps above and below the flat bands vanish, 
and only the $8$-band model continues to be trustworthy.
We find the $8$-band model can be reliably constructed down to $\theta = 0.6^\circ$, 
with only technical limitations of the projection technique restricting further reduction 
in twist angle.

The model construction is performed with a multi-step Wannier projection technique, 
which utilizes the chiral-symmetric \kp model \cite{CS_TWBLG} to define a proper 
subspace of the fully relaxed model.
In this context the chiral-symmetric model is not only an interesting theoretical limiting case, but also an extremely useful numerical tool.
The reduced models we obtain can serve as a springboard for studying correlated phenomena, but are a useful basis for performing other calculations that depend on 
accurate electronic structure.
We also study the response of these models under two common symmetry-breaking perturbations to the TBLG system: sublattice symmetry breaking due to the presence of a hexagonal boron nitride substrate and layer symmetry breaking due to an external electric displacement field.
Compared to previous efforts to produce a set of localized orbitals for 
the low-energy states of TBLG, the models presented here explicitly preserve 
all symmetries of the full tight binding models.
Furthermore, they are based on realistic \textit{ab-initio} tight binding band structures 
and avoid any empirical choices of coefficients or artificial band fitting procedures.

The inclusion of additional bands in our models provides clear advantages over existing treatments that focus on only the flat bands.
First, all bands within 150 meV of the charge neutrality point are retained near the magic angle, allowing for the treatment of an interaction strength which is comparable to the superlattice gap separating the nearly flat bands from the higher bands.
Secondly, all symmetries and band topology are manifest which can allow, for example, one to accurately describe how Chern bands emerge from the staggered chemical potential of a hexagonal boron nitride substrate.
Third, the inclusion of additional bands allows for the systematic study of angle dependence beyond the small range where the flat band manifold separates from all other bands.

Because of valley and spin symmetry, dealing with a 20 band ($5 \times 4$) or 32 band ($8 \times 4 $) model is challenging within any exact multi-particle numerical method.
However, simplifications could arise from the fact that most of the weight in the flat bands are from the triangular lattice sites within our final Wannier basis.
Since the additional bands could be well approximated as being completely filled or empty, it might be possible to supplement standard numerical methods with such constraints to reduce the computational effort required.
In any case, a useful middle ground must be found between completely ignoring the additional bands (which would give the wrong topology) and including them fully within a numerical calculation (which is prohibitively expensive).
This challenging step is left to future work.

\section{Methods}

\subsection{Wannierization Difficulties in TBLG}

Electronic bands with weak energy dispersions do not always imply trivial Bloch wave function texture throughout the Brillouin zone (BZ).
For example, the flat Landau levels formed in the presence of magnetic field feature non-trivial windings which lead to the topological Chern number~\cite{Berry_rev,Chern_TKNN}.
Such topological character is known to obstruct the construction of real space Wannier states~\cite{Vanderbilt_Chern}.
On the other hand, internal and crystalline symmetries can further constrain the Hilbert space structure for a group of bands, which can have consequences in obtaining Wannier orbitals in the real space.
For example, in the $Z_2$ topological insulator the group of nontrivial bands can only be described by a set of exponentially localized Wannier orbitals that break the time-reversal symmetry~\cite{Vanderbilt_Z2}.

In TBLG near the magic angle, it is relevant to construct the minimal tight-binding model as the basis for interacting theories.
The natural choice is to construct a model for just the group of flat bands.
For a single-valley model without spin polarization, these are the two bands closest to the Fermi level, and throughout we will refer to this two-band manifold as the ``flat bands'' even when considering twist angles far from the magic angle.
It has been shown that when all the (emergent) spatial symmetries are taken into account (with point group $D_6$)~\cite{Luijun_tBLG}, one cannot have a real space two-band representation of the flat bands~\cite{Adrian_Mott,Luijun_tBLG}.
This obstruction can be understood from the net chirality of two Dirac cones or from the symmetry representations of the flat bands~\cite{Adrian_Mott,Luijun_tBLG}.
One way to circumvent the obstruction is to give up on keeping all symmetries of the system manifest, such as by forgoing the $U(1)_v$ valley symmetry~\cite{Adrian_Mott} or the two-fold rotation symmetry~\cite{Liang_tBLG}.
These non-manifest symmetries are then realized in a nontrivial manner. 
The Wannier functions derived by reducing the symmetry form a honeycomb lattice and have a propeller-like (or "fidget spinner") shape, created by the superposition of three charge pockets at neighboring AA-stacking spots~\cite{Adrian_Mott,Liang_Wan,TBLG_wan,arxiv1905.01887}.
These models tend to have long hopping range and more complicated interaction terms due to the peculiar shape of their Wannier orbitals.
Another way to resolve the obstruction is to augment the flat band manifold with additional degrees of freedom which are artificial as in Refs.~\cite{Luijun_tBLG,Bernevig_TWBLG} which doubled the bands.
This new set of topologically non-trivial bands cancels the topology of the original flat bands.
However, these new degrees of freedom are not motivated physically from the realistic electronic structure and can complicate the modeling for interactions.

The other way to avoid the obstruction while retaining the symmetry is to include an additional group of bands from realistic electronic band structure to render the overall Hilbert space trivial in topological character.
In Ref. \onlinecite{Ashvin_10band}, it has been pointed out that the root for the obstruction in TBLG's flat bands is a ``fragile topology''~\cite{Adrian_fragile,Cano2018}.
This topological obstruction is relatively weak (fragile) when compared to the more robust Chern index, in the sense that the non-triviality can be canceled by including additional atomic insulator bands. 
%which are topological trivial.
The resulting Hilbert space can then be represented by exponentially localized Wannier orbitals.
This fragile topology resolution is not unique, and different solutions have been discussed which depend on the choice of additional bands. Requiring that the expanded space captures only nearby bands imposes a physical constraint that can be used to narrow this choice.
By paying the cost of additional tight-binding orbitals, these resolutions generally yield Wannier functions which are more conventional and short range as compared to the propeller solutions.

\subsection{Maximally Localized Wannier Functions}

A common approach to generate a localized model is to construct maximally localized Wannier functions (MLWF) \cite{Marzari1997, Marzari2012}.
They define tight-binding wavefunctions, $\phi_n(\vec{r})$, which minimize the total wavefunction spread

\begin{equation}
\Omega = \sum_n \left[ \bra{\phi_n} r^2 \ket{\phi_n} - (\bra{\phi_n} \vec{r} \ket{\phi_n})^2 \right].
\end{equation}

The simplest construction of MLWF for $N$ bands relies on the calculation of $M \geq N$ Bloch states $\psi_{m \vec{k}}$ over a uniform sampling of the BZ, and then evaluation of the overlap matrix elements of the periodic parts of the Bloch states, $u_m^{(\vec{k})}$ , 
\begin{equation}
M^{\vec{k},\vec{b}}_{mn} = \braket{u_m^{(\vec{k})}|u_n^{(\vec{k+b})} }
\end{equation}

where $\vec{b}$ is the momentum connecting each $\vec{k}$ to its nearest neighbor on the grid sampling of the BZ.
The overlap elements are then used to calculate a set of $\vec{k}$-dependent unitary rotations $U_{mn}^{(\vec{k})}$ that gives the proper mixing of the original Bloch states over the BZ to minimize the real space spread \cite{Marzari1997}.
Then the $n^\textrm{th}$ Wannier function at unitcell $\vec{R}$ is given by

\begin{equation} \label{eq:wannier}
    \ket{\vec{R} n} = \frac{V}{(2 \pi)^3} \int_{BZ} d\vec{k} e^{-i \vec{k} \cdot \vec{R}} \sum_{m=1}^N U_{mn}^{ (\vec{k}) } \ket{ \psi_{m\vec{k}} }
\end{equation}

Applying this methodology to a \kp model is not entirely trivial however.
If one attempts to construct a sampling of the $\vec{k}$-points of the BZ, one will find that the basis elements of the low-energy expansion are not smoothly periodic across the BZ boundary.
That is to say: the $n^\textrm{th}$ basis element of the continuum Hamiltonian at $\vec{k}$ will not correspond to the $n^\textrm{th}$ basis element at $\vec{k}+\vec{\delta}$ if $\vec{\delta}$ crosses a BZ boundary.
This is an unavoidable consequence of the non-periodic truncation of the low energy continuum model.
The problem can be alleviated by selecting only a subset of bands, and manually matching basis-elements that have nearly identical momentum.
This allows for an approximate evaluation of of $M^{\vec{k},\vec{b}}_{mn}$ even when $\vec{k+b}$ crosses a BZ boundary.

Even with these choices, the 8-band and 5-band electronic structures of TBLG are still not easily described with this approach.
An unconstrained optimization of the MLWF (using, for example, the package wannier90 \cite{Mostofi2008}) results in Wannier functions that have no clear symmetry within the moir\'e supercell, and the required band structure symmetries are impossible to impose as one truncates the range of the resulting tight-binding model.
Selectively localized Wannier functions (SLWF) \cite{Wang2014} can minimize the spread while constraining the Wannier function centers, but still cause similar symmetry issues without reducing the spread compared to our projection method.
There is a symmetry-constrained WF construction technique \cite{Sakuma2013} but currently it cannot be used alongside band disentanglement, a necessary step to properly construct the bands farthest from the Fermi energy.

The standard MLWF techniques perform poorly because while the wavefunctions of the flat band manifold are well localized in real space, the states just above and below the manifold are not.
When constructing a set of Wannier functions for both of these types of bands, the total spread functional $\Omega$ can be minimized by making all the Wannier functions of similar spread.
Doing so hybridizes the states and makes the interpretation of the WFs as a tight-binding model difficult.
In any case, the Wannier functions we obtain by projection have a possible improvement in total spread of less than $10\%$ (obtained by calculating both the gauge-dependent and gauge-independent parts of $\Omega$ \cite{Marzari1997}), meaning the projected orbitals are almost maximally localized anyway.

\subsection{Wannier Projection}

An alternate approach to constructing Wannier functions is by the projection of $M$ Bloch states $\psi_{m\vec{k}}(\vec{r})$ onto $N$ trial guesses $g_n(\vec{r})$ \cite{Marzari1997, Marzari2012, Mustafa2015}.
This is done by first defining the inner products
\begin{equation}
    a_{mn}^{ (\vec{k}) } = w_{m \vec{k}} \braket{\psi_{m \vec{k}} | g_n},
\end{equation}
where $w_{m \vec{k}} \in [0,1]$ is an optional weighting of the Bloch states, which can be tuned to allow for smooth band disentanglement.
Next, one computes $\tilde{a}$, the closest unitary approximation of $a$ (for now suppressing $\vec{k}$ dependence).
In practice, this is done by first factorizing $a$ through the singular value decomposition (SVD), $a = W \Sigma V^{\dagger}$, where $W, V$ are unitary matrices and $\Sigma$ is diagonal, and then setting 

\begin{equation}
    \tilde{a} = a \left[ V \frac{1}{\sqrt{ \Sigma^\dagger \Sigma}} V^\dagger \right]
\end{equation}

Note that this is equivalent to defining $\tilde{a} = a (a^\dagger a)^{-1/2}$.
Then, to obtain the WFs one simply replaces the matrix $U_{mn}^{(\vec{k})}$ in Eq. \ref{eq:wannier} with $\tilde{a}_{mn}^{(\vec{k})}$.

For this work, the choice of trial functions $g(\vec{r})$ and the weighting of the Bloch states $w_{m\vec{k}}$ are very important.
If the trial guess and the Hilbert subspace are not compatible enough, the SVD will be unable to create a good unitary approximation of the $a_{mn}^{(\vec{k})}$ matrix, resulting in poorly localized Wannier functions and failures in the tight-binding band structure.
The functions $g(\vec{r})$ can be initial guesses, or created iteratively by doing multiple Wannier projections of the same (or similar) Hamiltonians.
The weightings $w_{m\vec{k}}$ can depend on band-index or eigenvalue energy, or ignored if the number of selected bands is identical to the number of targeted Wannier functions.
We will later highlight how each of these choices of $g(\vec{r})$ and $w_{m \vec{k}}$ are used in various stages of our complete projection methodology.

\subsection{Choice of Initial Trial Functions}

Following Ref. \onlinecite{Ashvin_10band}, the 5-band and 8-band models must satisfy a representation matching equation to resolve the fragile topology of the flat bands.
To write down this equation, one must pick a lattice site and orbital symmetry for each Wannier function.
The candidate effective lattice sites are: (1) triangular, labeled $\tau$ and corresponding to $AA$ stacking regions, (2) honeycomb, labeled $\eta$ and corresponding to $AB/BA$ stacking regions, and (3) Kagome, labeled $\kappa$ and corresponding to intermediate stacking regions between $AB$ and $BA$ sites.
After considering relaxation of the TBLG system \cite{Dai2016, Nam2017, Zhang2018, Carr_GSFE, Yoo2019}, for $\theta < 1.5^\circ$ the honeycomb sites become large triangular domains of $AB$ or $BA$ stacking, and the Kagome lattice locations can be interpreted as domain-walls ($DW$) between them.

Along with these three sets of sites, we also consider four possible orbital symmetries, $s$, $p_z$, $p_+$ and $p_-$ ($p_\pm$ = $p_x \pm i p_y$).
These correspond to the same symmetry eigenvalues expected of a hydrogen orbital of the same label, for example the $s$ orbital is symmetric under $z \to -z$, but the $p_z$ orbital is anti-symmetric.
In the case of TBLG, the corresponding operation is physically implemented by a layer-exchanging two-fold rotation in three dimensional space about an axis running parallel to the layers.

With this groundwork covered, the representation matching equations can be summarized.
We use $(\textrm{f.b.})$ to represent the desired flat-bands.
For the 8-band model:
\begin{equation}
    (\tau, p_\pm) \oplus (\tau, s) \oplus (\eta,p_z) \oplus (\kappa,s) = (\kappa, s) \oplus (\kappa, p_z) \oplus (\textrm{f.b.})
\end{equation}
For the 5-band model including three bands above the flat-bands, which we call 5-band (top), the proper equation is:
\begin{equation}
   (\tau, p_\pm) \oplus (\tau, s) \oplus (\eta,p_z) = (\kappa, p_z) \oplus (\textrm{f.b.})
\end{equation}
and for the 5-band model including three bands below the flat-bands, which we call 5-band (bot), the proper equation is:
\begin{equation}
    (\tau, p_\pm) \oplus (\tau, p_z) \oplus (\eta,s) = (\kappa, s) \oplus (\textrm{f.b.})
\end{equation}

In every equation, the left-hand side corresponds to a realizable tight-binding model, while the right-hand side is a combination of topologically trivial states and the flat-bands with fragile topological obstruction.
The two 5-band models are related by swapping $p_z \leftrightarrow{} s$, and are defined by the $\Gamma$-point mirror-symmetry eigenvalues of the realistic bands just above or below the flat-bands.

These representation-matching equations prescribe the correct $g(\vec{r})$ trial guesses for a successful Wannier projection.
We build the guesses with a tight Gaussian wavepacket at the desired lattice site, and then assign the proper symmetry by hand by manipulating the phase of the layer and sublattice channels.
For example, to create a $p_z$ type orbital, one creates a Gaussian wavepacket with positive sign on the top layer and negative sign on the bottom layer.
These simple Gaussian trial functions change as they are projected into the Hilbert subspace of the low-energy band structure.
After successful projection, the resulting WFs have lattice centers and symmetry identical to their trail guesses, but they yield new interpretation in terms of stacking order and layer/sublattice character.
In doing so, we find that the symmetry-motivated prescription for each model can be connected to a physical ``moir\'e orbital'' motivated by the local stacking order.
The symmetries and interpretations of the WFs of each model are summarized in Table \ref{tab:orbs}.

\begin{table}
	\begin{tabular}{|c|c|c|r||c||c|}
		\hline
		\# & Lat. &  Sym. & Resulting orbital description & 5b top & 5b bot \\
		\hline
		1 & $\tau$   & $p_+$ & $B$ orbital dimers at $AA$           & $p_+$ & $p_+$\\
		2 & $\tau$   & $p_-$ & $A$ orbital dimers at $AA$           & $p_-$ & $p_-$\\
		3 & $\tau$   & $s$ & Equal mixture around $AA$              & $s$ & $p_z$\\
		4 & $\eta$   & $p_z$ & $A$ on $L_1$, $B$ on $L_2$ at $AB$   & $p_z$ & $s$\\
		5 & $\eta$   & $p_z$ & $B$ on $L_1$, $A$ on $L_2$ at $BA$   & $p_z$ & $s$\\
		6 & $\kappa$ & $s$   & $AB$ ($BA$) above (below)  $DW$      & $-$ & $-$\\
		7 & $\kappa$ & $s$   & $AB$ ($BA$) left (right) of  $DW$    & $-$ & $-$\\
		8 & $\kappa$ & $s$   & $BA$ ($AB$) left (right) of  $DW$    & $-$ & $-$\\
		\hline
	\end{tabular}
	\caption{
		Orbitals of the reduced Hamiltonians.
		The Wannier function index (\#), lattice description, and symmetry properties for the 8-band model are given in the first three columns, and describe both the initial trial Gaussian functions as well as the resulting Wannier functions.
		The symbols $\tau$, $\eta$ and $\kappa$ correspond to a triangular, hexagonal, and Kagome lattice, respectively.
		The symbols $p_\pm$, $p_z$, and $s$ refer to the symmetries of a hydrogenic wavefunction (see text for more details).
		The fourth column gives a description of the Wannier function character, including relations to stacking order and any layer or sublattice polarization.
		The last two columns list the symmetry content of the 5-band top and bottom models, with the rest of their Wannier functions' information identical to the 8-band case.}
	\label{tab:orbs}
\end{table}

\subsection{Multi-step Projection}\label{sec:multistep}

\begin{figure}
  \includegraphics[width=1\linewidth]{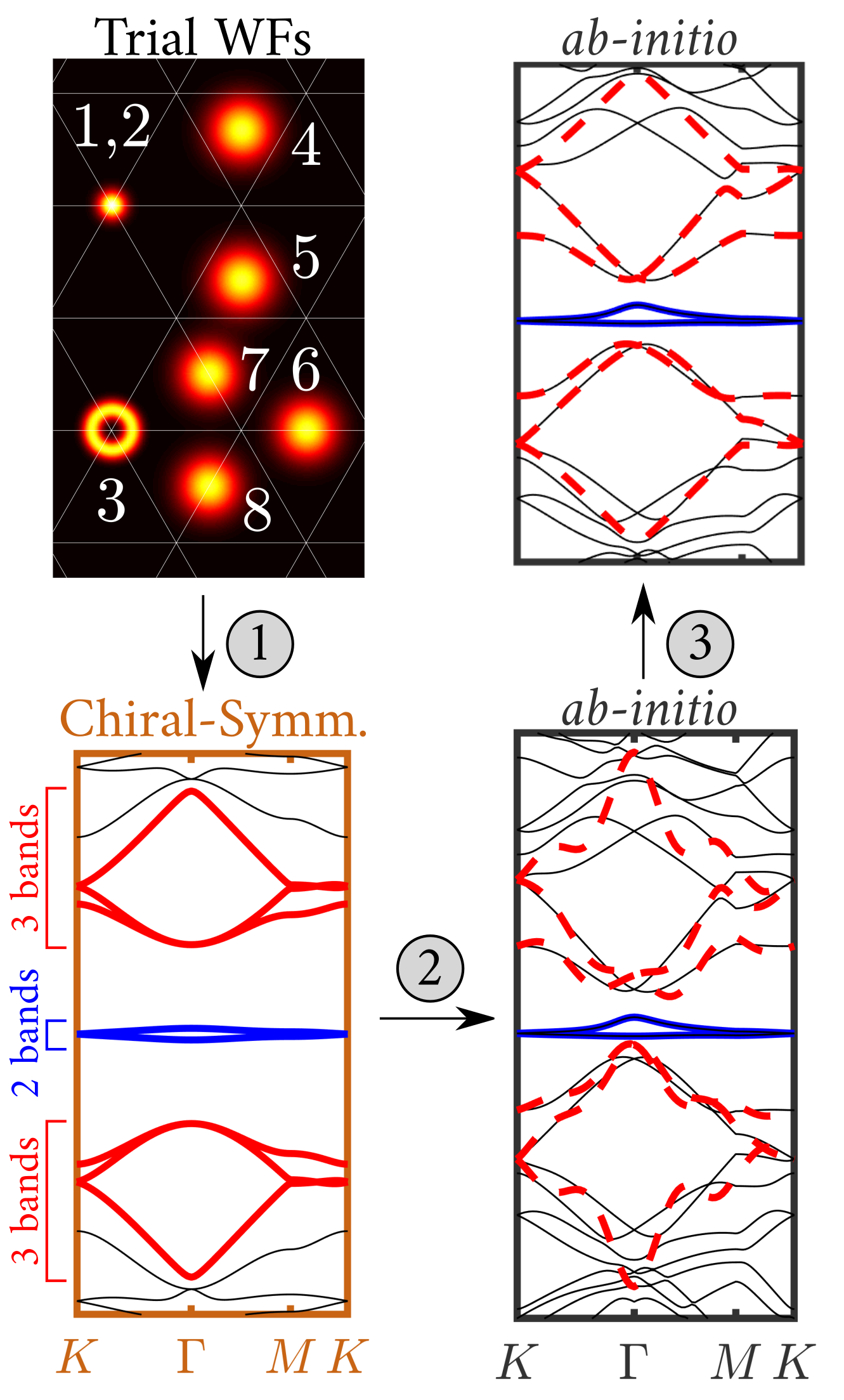}
  \caption{ 
  Overview of our projection methodology for the eight-band model, which is a combination of three iterative Wannierization steps.
  We begin with eight trial Wannier functions (WFs) following the symmetry prescription.
  These trials functions are projected into the chiral-symmetric model (Step 1), where a set two-band and three-band manifolds separate from the rest of the band structure.
  Combinations of the flat-bands (blue) and the three-band manifolds (red) can then  be projected into our \textit{ab-initio} continuum model (Step 2).
  The model is further improved with the use of an unbiased energy window for better band disentanglement (Step 3).
  The numbers correspond to the description of Sec.~\ref{sec:multistep}, and the model is further refined via the re-Wannierization explained in Sec.~\ref{sec:rewann} and Fig.~\ref{fig:rewan_example} before use.
  }
  \label{fig:methods_overview}
\end{figure}

The desired model will have perfectly accurate band structure for the flat-band manifold, with the accuracy of the band energies decreasing as one moves further away from the Fermi energy.
Our target Wannier functions therefore require a very careful projection of the low-energy band structure in order to controllably put most of the projection error in the higher-energy bands.
We remark that, as the higher-energy bands do not separate into well-defined groups defined by band gaps, certain arbitrariness will inevitably exist in how we disentangle these higher-energy bands. Here, we employ the projection technique \cite{Marzari2012} to select states that are smooth enough to allow for Wannierization.
Furthermore, in order to generate a full series of effective models over a large span of twist angles we eliminate all manual tuning in the process.
We achieve this automated construction with multiple Wannier projection steps, taking advantage of the chiral-symmetric limit \cite{CS_TWBLG} and improving the projection in an iterative fashion.
The entire process is schematically displayed in Fig. \ref{fig:methods_overview}.

\textbf{Step 1}: Use $N$ prescribed Gaussian trial guesses on the subspace defined by exactly $N$ bands of the chiral-symmetric (CS) model \cite{CS_TWBLG} to generate Wannier functions of the chiral symmetric model, $\textrm{WF}_\textrm{cs}$.
This is only possible in the CS model because the flat-bands and the three-band manifolds above and below the flat bands do not intersect with any other bands over the entire BZ.
Also perform a projection of only the three-band manifolds in the CS model to generate corresponding three-band WFs.

\textbf{Step 2}:
Use the three-band WFs as trial guesses for an energy-weighted projection of the realistic model, which disentangles a 3-band subspace from the multiple band crossings away from the flat-band manifold.
The projection is tuned depending on the angle, but always has full weight near the flat-bands, with exponentially decaying edges in the center of the three-band manifolds.
Combine the disentangled 3-band subspaces with the flat-band Bloch-states to generate a robust Hilbert subspace for the realistic model which still ensures perfect agreement along the flat-bands.
Use $\textrm{WF}_\textrm{CS}$ as trial guesses on the subspace just obtained to get the first Wannier functions for the realistic model, $\textrm{WF}_0$.
For angles near the magic angle (roughly $0.85^\circ$ to $1.2^\circ$), we find that $\textrm{WF}_0$ is already a very good model for the low-energy structure, but can have some band structure errors near the $\Gamma$ point.
% The following two steps help clean these problems, but these steps are mostly needed for when the gaps between the flat-band manifold and the trio-bands close in the realistic model.
The following step helps remove these problems, but is generally not necessary when there are sizable band gaps between the flat-band manifold and the three-band manifold above and below the flat bands.

\textbf{Step 3}: Use $\textrm{WF}_0$ as trial WFs on an unweighted subspace of the \textit{ab-initio} low-energy bands, usually defined by a hard energy cutoff.
This cleans up spurious band hybridizations at low angles for the 8-band model, and generally makes the band disentanglement more accurate for both the 8 and 5-band models.
We call these improved Wannier functions $\textrm{WF}_1$.

In general, all of these projections can be stored in $\vec{k}$-space or real space.
The $\vec{k}$-dependent $n \times n$ models are defined over a Brillouin  zone sampling of the moir\'e supercell, so one can easily integrate the result for different Wannier centers to generate an $n \times n$ real space model.
The $\vec{k}$-space models are sometimes difficult to transfer between the chiral-symmetric and realistic model, especially if one uses a different truncation radius for the two models (the CS model tends to converge much faster).
For this reason, the resulting Wannier projections are usually stored in real space on a uniform grid sampling of the moir\'e superlattice.
By sampling in this manner, one can in principle use the $\textrm{WF}_1$ generated at some angle $\theta$ as a trial guess for a one-shot projection at $\theta + \delta$.
We find this is not always a stable process, and instead use the full multistep projection process over a range of different angles.

As an aside, we note that the flat bands of twisted double bilayer graphene (TDBLG) \cite{Shen2019,Cao2019,Liu2019,Chebrolu2019,Koshino2019} are much simpler to Wannierize properly.
In TDBLG the $C_2$ symmetry is broken, which prevents its flat bands from entangling with the rest of the band structure, and thus one can accurately discuss them without invoking additional bands.
When there is no Chern number for the flat bands, a one orbital model for the active band can be constructed.
In the presence of a Chern number, a Haldane like model for the two opposite Chern number bands can instead be built.
This is a worthwhile exercise but relatively straightforward, and does not require the careful multi-step projection technique introduced here.

\subsection{Redistribution of Orbital Character}\label{sec:rewann} 

\begin{figure*}
  \includegraphics[width=1\linewidth]{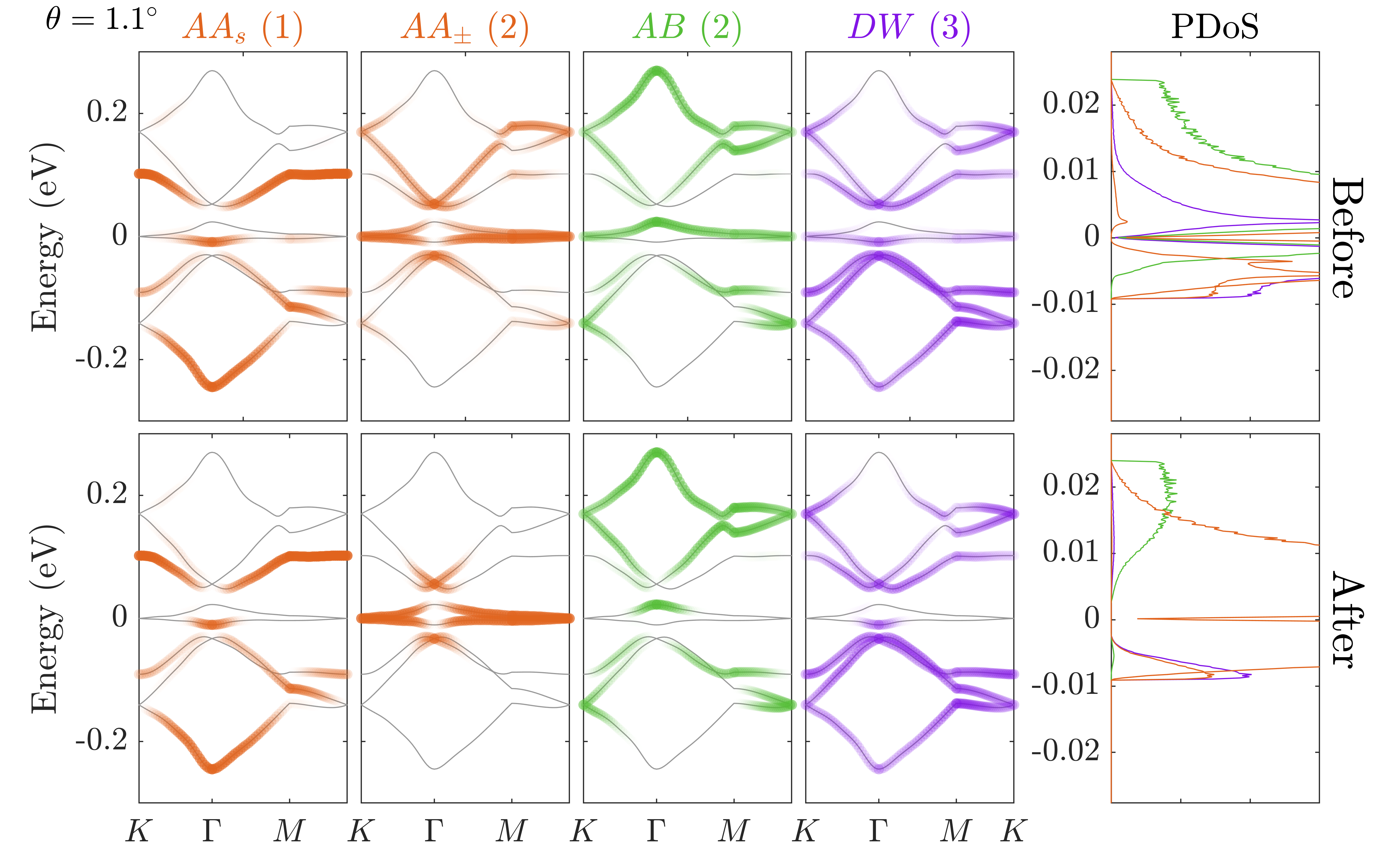}
  \caption{
  Projected band weightings for the Wannier functions of an 8-band model, both before (top) and after (bottom) a targeted re-Wannierization procedure.
  The goal of the re-Wannierization is to make the flat bands primarily $AA_\pm$ orbital character.
  The first four panels show the projected band weights in color for the four distinct types of Wannier functions, in order of increasing degeneracy: $AA_s$ (\#3), $AA_\pm$ (\#1,2), $AB$ (\#4,5), and $DW$ (\#6,7,8).
  The final panel shows the projected density of states (PDoS) near the Fermi-energy, highlighting the change in the orbital character of the flat bands.
  }
  \label{fig:rewan_example}
\end{figure*}

While the projection technique described allows us to construct reduced models with the desired properties, the definition of the effective orbitals still has some arbitrariness which allows further refinement.
More precisely, within the reduced Hilbert space one can still perform a basis rotation and redefine each of the effective orbitals (subjected to the symmetry constraints).
This should be contrasted with conventional solid-state systems, in which the notion of atomic orbitals is fixed by the microscopic lattice.

It is expected that the most important interaction terms for TBLG's flat bands arise due to the electron density that forms on the $AA$ stacking sites near the charge neutrality point.
Due to our symmetry prescription, these areas of high electron density can be best captured by the $AA_\pm$ orbitals in our models.
In anticipation of eventually incorporating electron-electron interaction, it is desirable to utilize the remaining freedom in basis transformation to maximize the weights of the $AA_\pm$ orbitals (WF \#1 and \#2) in the nearly flat bands.
Doing so can simplify interpretations of the many-body physics and, as a practical consideration, reduce the number of two-body and four-body integrals required to properly define the many-body model.

To achieve the desired weighting, we perform a final "re-Wannierization" of the reduced models. Note that this is done entirely within the reduced basis obtained from the previous subsection.
We take our initial guesses to be the original Wannier Function basis (so $g_n = e_n$, a basis element of $H$), and the subspace is the entire reduced band-structure.

For the sake of argument, imagine that the nearly flat bands have the symmetry and topology of the $AA_\pm$ orbitals.
If this was true, we should simply construct Wannier functions for the flat band manifold alone and relabel them as the new $AA_\pm$ orbitals.
In reality, there is a topological obstruction to such a construction, and by considering the symmetry representations we know that the overlap between the nearly flat bands and the $AA_\pm$ orbitals has to vanish at the $\Gamma$ point.
In other words, the weights of the $AA_\pm$ orbitals will necessarily have to be transferred to the higher energy bands at (and near) the $\Gamma$ point.
This motivates us to define a weighting of the Bloch states , $w_{m\vec{k}}$, which depend on both the band index and the momentum.
We let the weighting for the two flat bands be $1$ at all $\vec k$, but for the two bands above the flat bands, as well as the two below, we incorporate them into the projection with weights $w_{m\vec{k}} = e^{-(k/\sigma_k)^2}$, i.e., they are only incorporated in the vicinity of the $\Gamma$ point.
For every $\vec k$, we perform a two-orbital projection onto the subspace spanned by the original $AA_\pm$ orbitals, which allows us to construct a new set of Wannier functions of the same symmetry character.
The remaining WFs are then reconstructed by a  projection of the orthogonal subspace  onto the remaining orbitals.

An example of the re-Wannierization of the 8-band model is presented in Fig. \ref{fig:rewan_example}.
One can see in both the projected band structures and the projected density of states (PDoS) that the flat bands in the original reduced model ("Before") do not have a clear orbital character, except for certain high-symmetry points at which symmetry constraints are in play.
After the re-Wannierization ("After"), the flat bands are almost entirely $AA_\pm$ character in the new basis, except for the small "leakage" at the $\Gamma$ point necessitated by the symmetry constraints.
Likewise, the other Wannier functions only have a small weighting in the flat bands, confined near the $\Gamma$ point.
Overall, the flat bands go from 58\% $AA_\pm$ character to 92\% $AA_\pm$ character, a significant improvement that ensures the dominant interaction terms relevant to the correlated phenomena in the flat bands will involve mostly the $AA_\pm$ Wannier functions.
As the re-Wannierization is simply a change of basis within the reduced Hilbert space, it has no effect on the band structure. Furthermore, we find that it does not increase nor decrease the reduced model hopping range for suitable choices of the weighting parameter $\sigma_k$.
We note in passing that, in principle, if different orbital character was desired in various bands, the above procedure could be modified to create such a model, but we will not pursue this point in our reduced models.

\section{Results}

\begin{figure*}[htp]
  \includegraphics[width=1\linewidth]{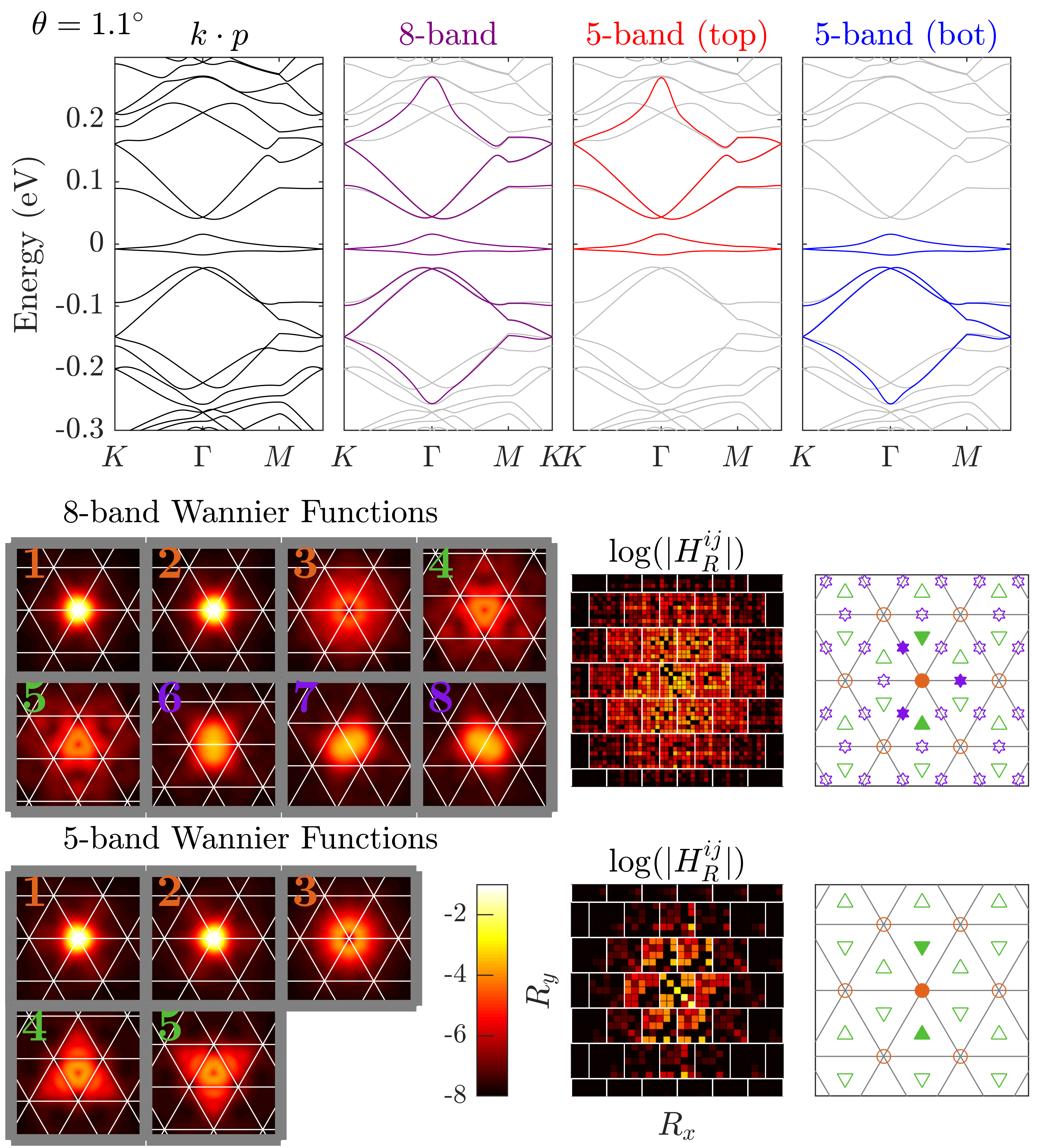}
  \caption{
  \textbf{Top}: Low energy band-structure at $\theta = 1.10^\circ$ for a realistic $\vec{k} \cdot \vec{p}$ model, 8-band model, and the two 5-band models.
  For the $n$-band models, the band structure near the flat-band manifold is identical to the $\vec{k} \cdot \vec{p}$ model (shown in faded grey).
  \textbf{Middle}: Overview of the 8-band model at $\theta = 1.1^\circ$.
  The amplitude of the Wannier functions, $|\Psi(r)|^2$, are shown in uniform color scale, and numbering with text color describing their lattice center (orange, green, and purple for the triangular, hexagonal, and Kagome lattice respectively).
  The superlattice moir\'e pattern is given by the thin white-lines, intersecting at $AA$ stacking areas.
  The middle panel shows the relative magnitudes of the hopping elements in the tight-binding Hamiltonian with a logarithmic color scale (from $10^{-1}$ to $10^{-8}$ eV).
  The white-lines divide $H_R$ for different moir\'e cells centered at $R$, and the $n \times n$ blocks represent the $(i,j)$ matrix element of $H^{ij}_R$.
  A schematic of the real-space tight-binding model is also provided on the right, with the centers of the Wannier functions in the fundamental unit-cell given by full color, while those at cells with $R \neq 0$ are given by colored outline.
  \textbf{Bottom}: Overview of the 5-band (top) model at $\theta = 1.1^\circ$.}
  \label{fig:results_overview}
\end{figure*}

\subsection{Reduced Tight Binding Hamiltonians}

An overview of the three models obtained from the multi-step projection and targeted re-Wannierization is shown for $\theta = 1.1^\circ$ in Fig. \ref{fig:results_overview}.
All three models perfectly reproduce the flat-band manifold of the \kp model, and also capture the relevant band-gap(s).
Even the band dispersion of the first two bands above and below the flat-band manifold are robustly captured.
Since these models are based on a single-valley in a \kp expansion, the underlying model lacks proper time-reversal symmetry, and as such the tight-binding Hamiltonians are generically complex.
Combining two copies of the Hamiltonian, one from the monolayer $K$ (already obtained) and one from $K'$ (which can be derived from $K$), would yield a tight-binding Hamiltonian that can be written with purely real coefficients after a linear transformation of the basis elements.
Such a model would have band structure identical to a fully atomistic tight-binding treatment, but naturally with twice as many orbitals as the single-valley model.
Approximating the inter-valley coupling to be zero, there is usually no reason to use the two-valley tight-binding model.
If inter-valley coupling is desired, it can be easily added in an empirical fashion to this two-valley model.

The Wannier functions have the proper symmetry, and most of their respective density remains close to their center (unlike the "fidget spinner" Wannier functions of the previous effective tight-binding models \cite{Adrian_Mott,Liang_Wan,TBLG_wan,arxiv1905.01887}).
The layer and sublattice projections of these functions is summarized in the last column of Table \ref{tab:orbs}.
The images of the wavefunctions in Fig. \ref{fig:results_overview} show only their magnitudes averaged over both layer and sublattice indices, but we have checked that they have underlying index polarization and complex phase consistent with their symmetry descriptions.
These interpretations of the model are similar to our previous and less-sophisticated treatment of a $10$-band model \cite{Carr_10band}.
We note that the interpretation of the projected density of states presented in the $10$-band study are related, but not identical, to the the electronic structure here.
In this model, the protected $s$ vs $p_z$ band-crossing of the flat-bands at the magic-angle still exists, but now one of the flat-bands at the $\Gamma$ point is a mixture of $AA$ and $DW$ ($s$) character, while the other is predominately $AB$ ($p_z$) character.
This is due to changes in the symmetry prescription between the $10$ and $8$ band models.

To accurately reproduce the flat-bands, the 5-band Hamiltonian needs to retain coupling terms only up to two moir\'e lengths ($2 \lambda_\theta$) and the 8-band needs only up to three ($3 \lambda_\theta$).
Thus for studying flat-band correlations the 5-band models are preferable, but for studying optical properties (or for correlations at twist angles well away from the magic-angle) the 8-band model is still usable.
Our approach differs from previous empirical treatment of a similar 5-band model (see Appendix of Ref. \onlinecite{Ashvin_10band}), as we obtain generic angle-dependence of all tight binding terms and couplings beyond nearest-neighbor.
Overall, these models are too complicated to give here in detail.
Instead, we provide tables of the values of the Hamiltonian as a function of twist-angle, and tools to generate the reduced models presented here from the realistic \kp model \cite{note:github}.

\subsection{Twist angle dependence}

\begin{figure*}[htp]
  \includegraphics[width=1\linewidth]{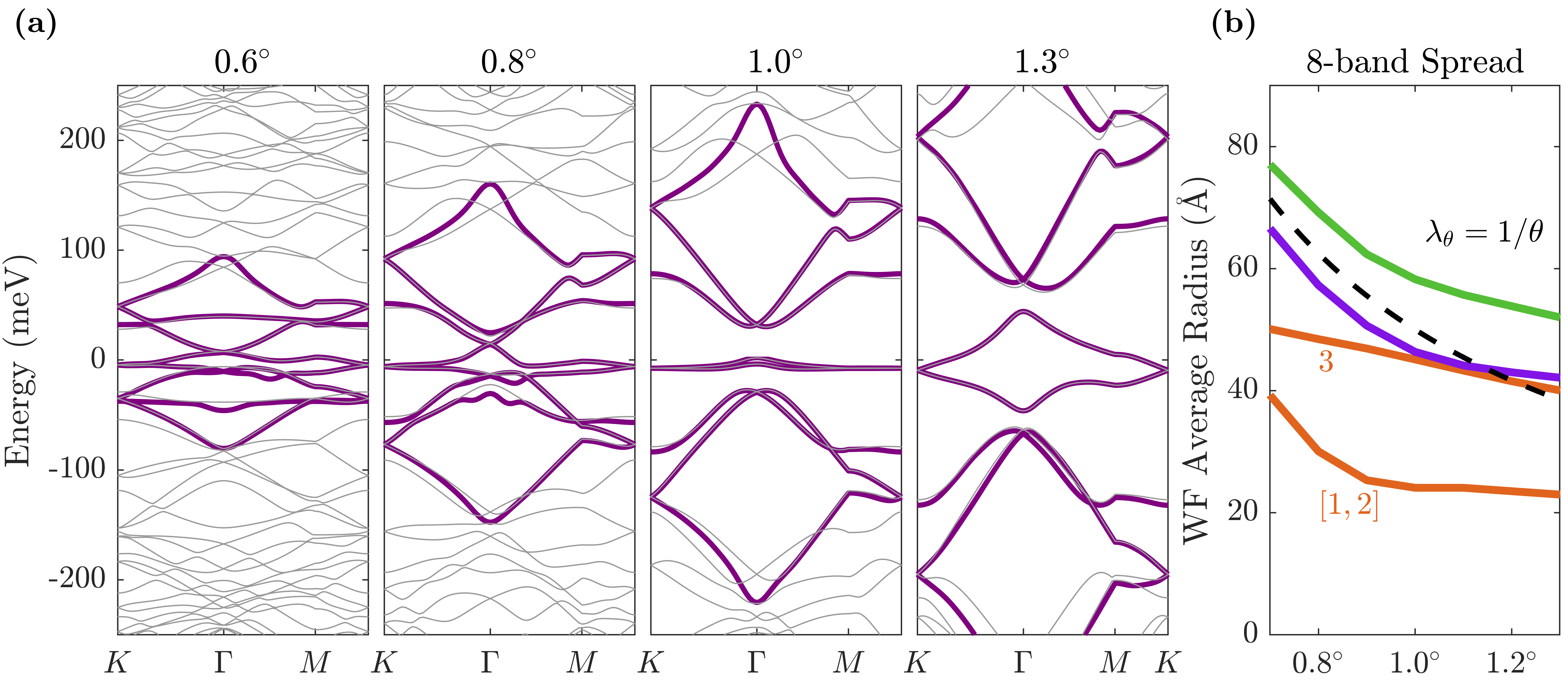}
  \caption{ 
            \textbf{(a)} Band structure comparisons between the 8-band model (purple) and the realistic $\vec{k} \cdot \vec{p}$ model (grey) for various twisting angles.
            \textbf{(b)} Angle dependence of the average radius of the 8-band model's Wannier functions, with colors matching the orbital numbering in Fig. \ref{fig:results_overview}.
            The radius is the the square root of that Wannier function's spread, $\Omega_n$.
            The radii of the $AB$ and $DW$ Wannier functions scale like the moir\'e length $\lambda_\theta$, while the size of the Wannier functions centered at $AA$ change much slower with $\theta$.
            The black dashed line showing $\lambda_{\theta} = 1/\theta$ is included for comparison.}
  \label{fig:theta_dep_fig}
\end{figure*}

The twist angle dependence of the 8-band model is summarized in Fig. \ref{fig:theta_dep_fig}.
We find that the automated multi-step projection technique is able to capture the low-energy band structure for $\theta \in [0.6^\circ, 1.3^\circ]$.
It fails outside of this angle range because of band intersections in the chiral-symmetric model.
For $ \theta > 1.3^\circ$, the three-band manifolds above and below the flat bands cross with the bands further from the Fermi energy (the thin black bands near $\Gamma$ in the left panel of Fig. \ref{fig:methods_overview}).
Similarly, for $\theta < 0.63^\circ$, the flat bands begin to mix with the the three-band manifolds above and below the flat bands, meaning the construction of seperate two-band and three-band manifolds is no longer well defined. 
The construction of Wannier functions past these twist angles is not impossible, but a projection technique different than the one we present here is necessary.

The dependence of the spread of the individual Wannier functions is also shown in Fig. \ref{fig:theta_dep_fig}b.
We see that the Wannier functions 1-3 (on the triangular lattice) do not scale like the moir\'e length $\lambda_\theta = 1/\theta$ near the magic angle.
This is because for $\theta \leq 1^\circ$, atomic relaxations cause the formation of $AA$ stacking regions that are fixed in size even as the angle decreases \cite{Dai2016, Nam2017, Zhang2018, Carr_GSFE, Yoo2019}.
The $AA$ stacking regions can be thought of as the intersection of the 1D strain solitons on the domain walls, and their size is determined through a balancing of in-plane strain energy to interlayer stacking energy \cite{Zhang2018}.
The other orbitals on the Kagome lattice (domain walls) and Hexagonal lattice ($AB$/$BA$) increase in spread as the twisting-angle decreases, which is not surprising as they are pinned to the geometry of the expanding domains and domain-walls.
When the gap between the flat band manifold and the nearby band manifolds closes ($\theta < 0.85^\circ$) we see Wannier functions \#1 and \#2 ($AA_\pm$) begin to grow in size, due to increasing hybridization with electronic structures not associated with the $AA$ stacking spots.

\subsection{Representing symmetry lowering perturbations in the reduced model}

\begin{figure}[htp]
  \includegraphics[width=1\linewidth]{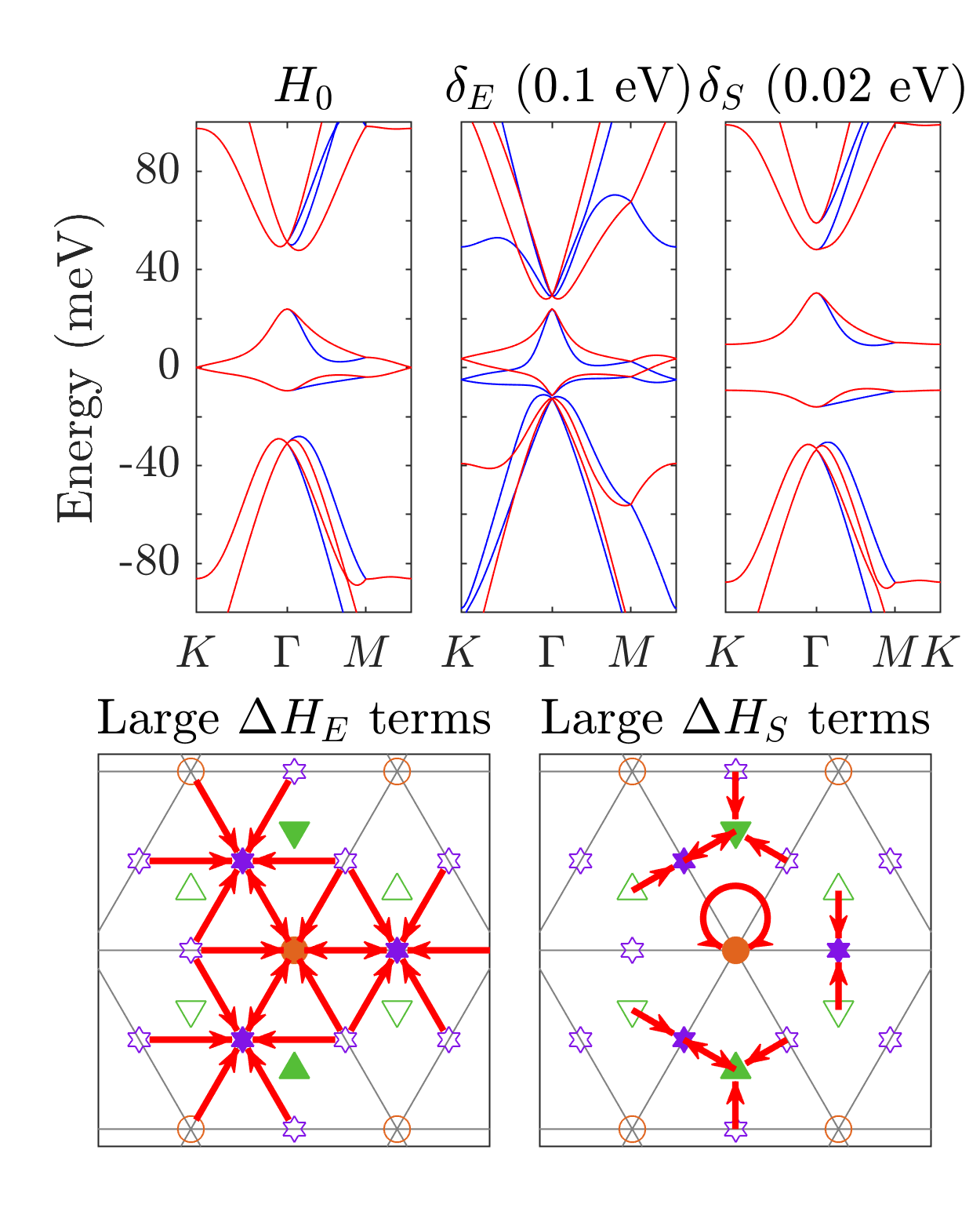}
  \caption{
    \textbf{Top}:
    Band structures of the $\vec{k} \cdot \vec{p}$ model at $\theta = 1.1^\circ$ with and without ($H_0$) the external perturbations of displacement field ($\delta_E$) and sublattice onsite energy ($\delta_S$).
    $\delta_S$ refers only to the layer-symmetric sublattice energy (e.g. the $A$ orbitals of both layers have positive energy, while the $B$ orbitals of both layers have negative energy).
    The red bands correspond to an expansion about the $K$ valley of the monolayer Brillouin zone, while the blue lines correspond to the $K'$ valley (a supercell tight-binding model will naturally include both).
    \textbf{Bottom}:
    Geometric schematic showing the largest terms in the perturbing Hamiltonians $\Delta H_i$, based on the structures introduced in Fig. \ref{fig:results_overview} for the 8-band model.
    The red-arrows correspond to large matrix elements that ``hop'' into a Wannier function of the fundamental moir\'e cell.
    The circular arrow on $\Delta H_S$ represents a large onsite term for Wannier functions $1$ and $2$ ($AA_\pm$), but not $3$ ($AA_s$).
    }
  \label{fig:perturb_fig}
\end{figure}

Two of the important symmetries of bilayer graphene system can be easily broken in experimental devices: the layer symmetry and sublattice symmetry.
Layer symmetry can be lifted by applying an external vertical electric displacement field.
This places the top and bottom layers at different electrostatic potentials, and to first order can be included in a \kp model by adding an onsite energy difference $\pm \delta_E$ for the top and bottom layer degrees of freedom (determined by the magnitude and sign of $E$).

The sublattice symmetry can be broken experimentally when the bilayer is encapsulated by nearly-aligned hexagonal boron nitride (hBN). 
hBN  has a similar crystal structure to graphene, but is composed of boron and nitrogen atoms instead of only carbon.
When graphene is in close proximity of aligned hBN,  sublattice symmetry is broken as a carbon atom near boron feels a slightly different electrostatic potential than for the one near nitrogen.
This can be similarly included in a \kp model by including a $\pm \delta_S$ for the $A$ and $B$ sublattice degrees of freedom (determined by $\alpha$ or $\beta$ alignment with hBN).
In general, this energy difference does not have to be the same for the two layers, as physically the top and bottom hBN encapsulations do not have to be simultaneously aligned in the same way.
To study any combination, we consider both layer symmetric sublattice energy, $\delta_S$, and layer anti-symmetric sublattice energy, $\delta'_S$.
For example, the layer symmetric perturbation means the $A$ ($B$) sublattice of both layers feels a positive (negative) onsite perturbation, while the layer anti-symmetric means the $A$ ($B$) sublattice of layer 1 is positive (negative) while the $A$ ($B$) sublattice of layer 2 is negative (positive).

For both types of perturbations, the new \kp Hamiltonian is given by $H'_k = H_k^0 + \delta_i \Delta H_k$.
One can easily calculate the first order perturbation  $\Delta H$ for the reduced Wannier models by taking the $n$-band subspace projector $P_n$ and sandwiching it around the perturbation of $H'_k$: $\Delta H = P_n^\dagger \Delta H_k P_n$.
The \kp band structures and a summary of the largest terms of $\Delta H_i$ are presented in Fig. \ref{fig:perturb_fig}.

If the two graphene layers were decoupled, the only effect of electric displacement field would be to move the Dirac cones of each layer away from one another in energy.
This can be seen in the \kp band structures as a valley-dependent energy shift of the states at the moir{\'e} $K$ points (the red and blue bands are no longer identical near $K$), which corresponds to a breaking of the layer-exchanging symmetries.
The 8-band reduced model captures this effect by having mirror-symmetry breaking terms added between the triangular and Kagome lattice orbitals.

For a graphene monolayer, the substrate symmetry breaking term opens up a finite gap at the Fermi energy, splitting the Dirac cone into two parabolas.
In the bilayer system, the gap remains and is roughly equal in magnitude to the value of $\delta_S$.
In the reduced 8-band Hamiltonian, the layer-symmetric sublattice perturbation manifests as ideal onsite energies for Wannier functions $1$ and $2$ ($0.5 \delta_S$ and $-0.5 \delta_S$, respectively), as well as some smaller coupling terms between the honeycomb and Kagome sites.
This is not surprising, as WFs $1$ and $2$ have the highest amount of sublattice polarization, while WF $3$ has the least.
We remark that, although the Dirac cones are gapped by an onsite energy difference, the resulting non-degenerate band (in each valley) carries a nonzero Chern number. This is a direct consequence of the topological character of the flat bands in TBLG \cite{Adrian_Mott}.
The anti-symmetric perturbation $\delta'_S$ is not shown in Fig. \ref{fig:perturb_fig}, but it has a similar band structure to $\delta_S$ with a smaller gap-opening (roughly $30\%$ of that for $\delta_S$).
The largest terms are couplings between orbitals $1, 2$ and $6, 7, 8$ (e.g. $AA$ to $DW$ couplings), but they have a smaller overall value of $0.15 \delta'_S$.

\section{Conclusion}

We have presented a robust projection scheme for obtaining accurate Wannier functions in twisted bilayer graphene.
The construction of these Wannier functions takes advantage of the chiral-symmetric limit of TBLG's low energy model, and can create Hamiltonians that reproduce realistic band structures even outside the magic angle regime.
Although we did not maximize the localization of these Wannier functions, their spread is close to the smallest possible for the chosen Hilbert subspace.

The Wannier functions exist on a triangular, honeycomb, and Kagome lattice.
This yields a direct interpretation in terms of local stacking order, allowing one to ascribe features of the electronic structure to interactions between effective orbitals defined over the moir\'e superlattice.
The low-energy Hamiltonians can be truncated symmetrically, and they need hoppings with lengths only up to 3 moir\'e lengths to accurately reproduce the flat bands near the magic angle.
Studying the Wannier functions as the twist angle changes shows that the functions defined on the $AA$ stacking (orbitals 1-3) do not scale like the moir\'e length $\lambda_\theta$ while those defined on the $AB$ or $DW$ geometry scale like $\lambda_\theta$, consistent with the features of atomistic relaxation in the system.
The effect of external electric displacement field and sublattice symmetry breaking cause clear changes to the low-energy band structure, and these changes are captured in the reduced model by changing different subsets of hopping terms.

The models presented here can be used as a starting point for unbiased study of correlated phases in TBLG.
Our models differ from previous approaches primarily in the resolution of TBLG's fragile topology: instead of breaking a symmetry we include additional bands outside of the flat band manifold.
This allows for the comparison of symmetry and symmetry-broken states and removes the unconventional long-range coupling of the propeller orbitals.
Furthermore, with complete freedom in the choice of twist angle, one can illuminate the nature of correlations near the magic angle and also predict the presence of other (similar or dissimilar) correlated phases in the range $[0.6^\circ, 1.3^\circ]$.
We leave such study to future work.
The reduced models are made publicly available as part of a larger suite of continuum models for relaxed TBLG \cite{note:github}.
These codes include the 8-band and 5-band Hamiltonians for various angles highlighted in this work and also allows for the construction of additional models with arbitrary twist angle and external perturbations.

\begin{acknowledgments}
HCP and AV thank L. Zou and T. Senthil for earlier related collaborations on TBLG. 
This work was supported by the STC Center for Integrated Quantum Materials, NSF Grant No. DMR-1231319 and by ARO MURI Award W911NF-14-0247. HCP was supported by a Pappalardo Fellowship at MIT and a Croucher Foundation Fellowship.
AV was supported by a Simons Investigator award and NSF-DMR 1411343.
\end{acknowledgments}

\vfill

\bibliography{wannier_tblg}

\end{document}